\documentclass[a4paper]{jpconf}

\usepackage{iopams}

\newcommand{\lanln}[1]{{\it Preprint\/} #1}
\newcommand{\BbbR}{\mathbb{R}}

\newcommand{\be}{\begin{eqnarray}}
\newcommand{\ee}{\end{eqnarray}}

\begin{document}
\title{Excited by a quantum field: Does shape matter?\footnote{Based
on a talk given at ``Recent Developments in Gravity" 
(NEB XII), Nafplio, Greece, 29 June -- 2 July 2006.}}

\author{Jorma Louko and Alejandro Satz}

\address{School of Mathematical Sciences, 
University of Nottingham, 
Nottingham NG7 2RD, 
United Kingdom}

\ead{jorma.louko@nottingham.ac.uk, pmxas3@nottingham.ac.uk\\[0.5cm] 
arXiv:gr-qc/0609062}

\begin{abstract}
The instantaneous transition rate of an arbitrarily accelerated
Unruh-DeWitt particle detector on four-dimensional Minkowski space is
ill defined without regularisation. We show that Schlicht's
regularisation as the zero-size limit of a Lorentz-function spatial
profile yields a manifestly well-defined transition rate with
physically reasonable asymptotic properties. In the special case of
stationary trajectories, including uniform acceleration, we recover
the results that have been previously obtained by a regularisation that
relies on the stationarity. Finally, we discuss evidence for the
conjecture that the zero-size limit of the transition rate is
independent of the detector profile.
\end{abstract}

\section{Introduction}

Starting with the seminal work of Unruh~\cite{unruh}, it has now been
recognised for 30 years that a uniformly accelerated observer in
Minkowski space sees Minkowski vacuum as a thermal state in
temperature $T = a/(2\pi)$, where $a$ is the magnitude of the proper
acceleration. This result is of interest already in its own right
within flat spacetime quantum field theory, and it has been confirmed
by a number of
methods~\cite{unruh,deWitt,byd,haag-book,wald-smallbook}. For
relativists, the result is of particular interest because of its close
mathematical similarity to the thermal properties of quantum fields in
stationary black hole spacetimes~\cite{unruh,hawking}.

A conceptually concrete way to address quantum effects in accelerated
motion is to analyse a particle detector coupled to the
quantum field. For the uniformly accelerated motion, a subtlety in
such an analysis arises from the fact that the motion is
\emph{stationary\/}, that is, the orbit of a timelike Killing vector. 
Because of stationarity, the first-order perturbation theory transition
probability over the whole trajectory is infinite, owing to the
infinite total proper time. However, this probability can be formally
factorised into the product of the total proper time and a finite
remainder, and the remainder can by stationarity be interpreted as the
transition rate per unit proper time
\cite{letaw-pfautsch,letaw,rosubookchapter,rosu}. This regularisation
prescription can be extended from stationary trajectories in Minkowski
space to curved spacetime~\cite{sonego-westman}, both for the
Unruh-DeWitt monopole detector \cite{unruh,deWitt} and a variety of
its generalisations. A~recent review can be found
in~\cite{Langlois-thesis}.

For nonstationary motions the transition rate need not be constant
along the detector's trajectory, and a regularisation that relies on
stationarity is no longer available. The first message of this talk
is:
\begin{itemize}
\item
For an Unruh-DeWitt monopole detector, the instantaneous transition
rate is ill-defined without regularisation.
\end{itemize}
This observation appears to have been first made by
Schlicht~\cite{Schlicht:thesis,schlicht}, who showed that a
conventional $i\epsilon$ regularisation yields a Lorentz-noninvariant
transition rate for uniformly accelerated motion. We discuss the
mathematical reason for this phenomenon and show that
the $i\epsilon$ regularisation leads to a Lorentz-noninvariant
transition rate for
\emph{every\/} non-inertial trajectory. The conventional $i\epsilon$
prescription 
does therefore not provide a physically acceptable regularisation for
the instantaneous transition rate. 

Schlicht proposed to regularise the transition rate in arbitrary
motion by making the detector spatially extended in its instantaneous
rest frame, with a spatial sensitivity profile that has a certain
fixed shape but depends on a size parameter, and then letting the size
parameter approach zero~\cite{Schlicht:thesis,schlicht}. He showed
that this regularisation yields the expected Planckian spectrum for
uniform acceleration, and he analysed a selection of nonstationary
trajectories via mainly numerical methods. We show that Schlicht's
regularisation yields a well-defined transition rate for every
trajectory satisfying certain technical conditions, and we express the
result as a manifestly finite integral formula that no longer involves
regulators or limits. For the stationary trajectories the result
agrees with that obtained in \cite{deWitt,letaw-pfautsch,letaw}, 
and for nonstationary trajectories we extract asymptotic
results that appear physically reasonable. The second message of this
talk thus is:
\begin{itemize}
\item
A spatial sensitivity profile is a viable regulator for the
instantaneous transition rate. 
\end{itemize}

The rest of the talk will put some flesh on these messages. The main
conclusions rely on a particular choice of the spatial profile
function, but in section \ref{sec:shape?}  we present some evidence
suggesting that that the zero size limit may be insensitive to the
detailed form of the profile. The talk is based on~\cite{louko-satz},
where further detail can be found.

We work in four-dimensional Minkowski spacetime with metric signature
$({-} {+} {+} {+})$ and in units in which $\hbar= c=1$. Boldface
letters denote spatial three-vectors and sans-serif letters spacetime
four-vectors, and a square of a spatial vector (respectively spacetime
vector) is understood in the sense of the Euclidean (Minkowskian)
scalar product.

\section{Unruh-DeWitt detector}
\label{sec:detectormodels}

Consider a pointlike detector that moves in four-dimensional Minkowski
space on the world line~$\mathsf{x}(\tau)$, where $\tau$ is the proper
time. We take the detector to have two quantum states, denoted by
$\vert0\rangle_d$ and $\vert1\rangle_d$, which are eigenstates of the
detector internal Hamiltonian $H_d$ with 
the respective eigenvalues $0$
and $\omega$, $\omega\ne0$. The detector is coupled to the real,
massless scalar field $\phi$ with 
the 
interaction Hamiltonian
\begin{equation}
H_{\mathrm{int}}=c \chi(\tau)
\mu(\tau)\phi
\bigl(\mathsf{x}(\tau) \bigr)\ , 
\label{eq:Hint} 
\end{equation}
where $c$ is a coupling constant
and 
$\mu(\tau)$ is the detector's monopole moment operator, evolving in
the Heisenberg picture under~$H_d$. 
$\chi(\tau)$ is a switching function, which specifies how the
interaction is switched on and off by an external agent. 

Suppose first that the switching function is smooth and has
compact support, so that the 
initial and final states can be described in
terms of the uncoupled system. If before the interaction
the field is in the Minkowski vacuum $\vert 0\rangle$ and the detector
in the state~$\vert0\rangle_d$, the first-order perturbation theory
probability of finding the detector in the state $\vert1\rangle_d$
after the interaction is
\cite{unruh,deWitt,byd,wald-smallbook,Junker}
\begin{equation}
\label{eq:total-probability}
P(\omega)
=
c^2
\, 
{\bigl\vert
{}_d\langle0\vert\mu(0)\vert1\rangle_d\bigr\vert}^2
F(\omega)
\ , 
\end{equation}
where the response function 
$F(\omega)$ is given by 
\begin{equation}
\label{defresponse}
F(\omega)= \int_{-\infty}^{\infty}
\mathrm{d}\tau' \int_{-\infty}^{\infty}\mathrm{d}\tau'' 
\, 
\mathrm{e}^{-i\omega(\tau'-\tau'')}
\, 
\chi(\tau')\chi(\tau'')
\, 
W(\tau', \tau'') 
\ , 
\end{equation}
and the correlation function 
$W$ is the pull-back of the Wightman distribution, 
\begin{equation}
\label{eq:W} 
W(\tau',\tau'') := 
\langle 0\vert 
\phi \bigl(\mathsf{x}(\tau') \bigr)
\phi \bigl(\mathsf{x}(\tau'') \bigr)
\vert 0\rangle
\ . 
\end{equation}
As $W$ 
is a well-defined distribution on 
$\BbbR \times \BbbR$ \cite{Junker}, 
the transition probability given by 
(\ref{eq:total-probability})--(\ref{eq:W}) is well defined. 

Suppose then that 
no friendly neighbourhood external agent 
is available to switch 
the interaction off before we observe the detector. We wish to ask: 
\emph{What is the probability of finding the detector in
the state $\vert1\rangle_d$
while the interaction is still switched on?\/} 
This is arguably the question encountered in a practical
measurement where one looks at an ensemble of accelerated detectors
(say, atoms or ions) at a given moment of time and counts what
fraction of the detectors are in an excited state. 

An attempt to answer this question within the 
detector model (\ref{eq:Hint}) would be to introduce
in the switching function a sharp cutoff, $\chi(\tau') \to \chi(\tau')
\Theta(\tau-\tau')$, where $\tau$ is the proper time at which the
detector is read and $\Theta$ is the Heaviside function. If we further
push the switch-on to the asymptotic past, this would mean making in
(\ref{eq:total-probability})--(\ref{eq:W}) the replacement
$\chi(\tau') \to
\Theta(\tau-\tau')$. Formal manipulations then yield for the
$\tau$-derivative of the response function the expression
\begin{equation}
\label{defexcitation-sharp-infty}
\dot{F}_{\tau} (\omega)
=
2 \, 
\mathrm{Re}
\int_{0}^{\infty}
\mathrm{d}s
\,\,
\mathrm{e}^{-i\omega s}
\, 
W(\tau,\tau-s)
\ . 
\end{equation} 
$\dot{F}_{\tau} (\omega)$ differs from the instantaneous transition
rate only by a multiplicative constant that is independent of the
trajectory, and we shall from now on suppress this constant.

The problem with these manipulations is that formula 
(\ref{defexcitation-sharp-infty})
is ambiguous, 
since $W$ is a distribution with a singularity at the coincidence limit
and the integration range
has a sharp boundary at this
singularity. To see that the problem is
significant, suppose we go to a specific Lorentz frame, $\mathsf{x} =
(t, \mathbf{x})$, replace 
the Wightman distribution by its
conventional $i\epsilon$ regularisation, 
\begin{equation}
\label{tradWightman-eps}
\langle 0\vert \phi(\mathsf{x})\phi(\mathsf{x}')\vert 0\rangle_\epsilon
=
\frac{-1}{4\pi^2}\frac{1}{{(t-t'-i\epsilon)}^2
- 
{\vert\mathbf{x}-\mathbf{x'}\vert}^2}
\ , 
\ \ 
\epsilon >0 \ , 
\end{equation}
and take the limit $\epsilon \to 0_+$ after performing the integral in
(\ref{defexcitation-sharp-infty}). 
Assuming that the trajectory is sufficiently differentiable and has 
suitable falloff properties in the distant past, the result is 
\cite{louko-satz} 
\begin{eqnarray}
\dot{F}_{\tau}(\omega)
&=&
-\frac{\omega}{4\pi}+\frac{1}{2\pi^2}
\int_0^{\infty}\textrm{d}s
\left( 
\frac{\cos (\omega s)}{{(\Delta \mathsf{x})}^2} 
+ 
\frac{1}{s^2} 
\right) 
\nonumber
\\
&& 
-\frac{1}{4\pi^2}\frac{\ddot{t}}{{(\dot{t}^2-1)}^{3/2}}
\left[ \dot{t}\sqrt{\dot{t}^2-1}
+\ln\!\left(\dot{t}-\sqrt{\dot{t}^2-1}\,\right)\right] 
\ , 
\label{resultado1-noninvariant}
\end{eqnarray}
where $\Delta \mathsf{x} := \mathsf{x}(\tau) - \mathsf{x}(\tau-s)$. 
The last term in (\ref{resultado1-noninvariant}) 
vanishes for inertial
trajectories but is Lorentz-noninvariant wherever the proper
acceleration is nonzero. In the usual distributional setting of integrating against smooth test functions, the
functions (\ref{tradWightman-eps}) duly converge to the Lorentz-invariant
Wightman distribution as $\epsilon\to0$~\cite{kay-wald}, 
but the instantaneous transition
rate (\ref{defexcitation-sharp-infty}) falls outside this setting because of the sharp switch-off 
and retains a Lorentz-noninvariant piece even 
in the limit
$\epsilon \to0_+$. 

\textbf{Moral}: The instantaneous transition rate
(\ref{defexcitation-sharp-infty}) is ill-defined as it stands and 
needs to be regularised.

\section{Spatial profile}
\label{sec:profile}

Schlicht \cite{Schlicht:thesis,schlicht} proposed to regularise the
transition rate (\ref{defexcitation-sharp-infty}) by giving the
detector a spatial sensitivity profile that is rigid in the detector's
instantaneous rest frame. 
This idea can be motivated by the
observation that real material systems 
(say, atoms or ions) are not
pointlike. 


Technically, Schlicht's proposal is to replace the field operator in
the interaction Hamiltonian (\ref{eq:Hint}) by a spatially smeared
field operator,
\begin{equation}
\label{eq:smeared-phi}
\phi\bigl( \mathsf{x}(\tau) \bigr) 
\to 
\int \mathrm{d}^3 \xi
\,
\epsilon^{-3} f (\boldsymbol{\xi}/\epsilon)
\, 
\phi 
\left( \mathsf{x}(\tau)  +\xi^i\mathsf{e}_i(\tau)  \right)
\ , 
\end{equation}
where $\mathsf{e}_i$ 
are three unit vectors that together with the
velocity $\dot{\mathsf{x}}$ 
form an orthonormal tetrad, 
Fermi-Walker transported along the trajectory. 
The four quantities $(\tau,\boldsymbol{\xi})=(\tau,\xi^1,\xi^2,\xi^3)$ 
are thus 
Fermi-Walker coordinates in a neighbourhood of the trajectory~\cite{mtw}. 
The profile function $f: \BbbR^3 \to \BbbR$ is assumed to be
non-negative and to integrate to unity, and $\epsilon$ is a positive
parameter that determines the characteristic size of the smeared
detector. When $f$ is chosen to be the Lorentzian function, 
\begin{equation}
\label{lorentzian}
f(\boldsymbol{\xi})
=
\frac{1}{\pi^2}
\frac{1}{\bigl({\vert\boldsymbol{\xi}\vert}^2+1\bigr)^2}
\ , 
\end{equation}
Schlicht 
showed that $W$ in
(\ref{defexcitation-sharp-infty}) gets replaced by 
\begin{equation}
\label{corrSchlicht}
W_\epsilon (\tau,\tau')
=
\frac{1}{4\pi^2}\frac{1}{\left[ \mathsf{x}-\mathsf{x}'
-i\epsilon 
(\dot{\mathsf{x}}+\dot{\mathsf{x}}') 
\right]^2}
\ , 
\end{equation}
where the unprimed and primed quantities
are evaluated respectively at $\tau$ and~$\tau'$. 
Note that $W_\epsilon$ (\ref{corrSchlicht}) 
is manifestly Lorentz covariant. 
Schlicht further showed that 
the $\epsilon\to0_+$ limit yields the Planckian spectrum 
for the uniformly accelerated motion, 
thus agreeing with the regularisation that 
relies on stationarity 
\cite{unruh,byd,wald-smallbook}. 
He also examined the $\epsilon\to0_+$ limit for a 
number of other trajectories, with physically reasonable results. 

Schlicht's results have been generalised by P.~Langlois
\cite{Langlois-thesis,Langlois} to a variety of situations, including
Minkowski space in an arbitrary number of dimensions, quotients of
Minkowski space under discrete isometry groups, the massive scalar
field, the massless Dirac field and certain curved
spacetimes. Langlois also observed that an alternative way to arrive
at $W_\epsilon$ (\ref{corrSchlicht}) is to regularise the mode sum
expression for the Wightman function by an exponential frequency
cut-off in the detector's instantaneous rest frame, rather than in a
fixed Lorentz frame.

\section{Lorentz-function profile: Zero-size limit}
\label{sec:zero-size-limit}

When the regularised correlation function (\ref{corrSchlicht}) is
substituted in~(\ref{defexcitation-sharp-infty}), the existence of an
$\epsilon \to 0_+$ limit is not obvious for an arbitrary trajectory
since $\epsilon$ appears under the integral. However, for
trajectories that are sufficiently differentiable and have suitable
falloff properties in the distant past, the limit exists and equals
\cite{louko-satz}
\begin{equation}
\dot{F}_{\tau}(\omega)
= 
-\frac{\omega}{4\pi}+\frac{1}{2\pi^2}
\int_0^{\infty}\textrm{d}s
\left( 
\frac{\cos (\omega s)}{{(\Delta \mathsf{x})}^2} 
+ 
\frac{1}{s^2} 
\right) 
\ . 
\label{resultado1-infty}
\end{equation}
Since the integrand in (\ref{resultado1-infty}) remains finite at
$s\to0_+$ and since ${(\Delta \mathsf{x})}^2 \le -s^2$, formula
(\ref{resultado1-infty}) is manifestly well-defined.

Formula (\ref{resultado1-infty}) gives the transition rate as split
into its odd and even parts in~$\omega$. Another useful split is into
the inertial part and the noninertial correction, as introduced for
stationary trajectories in \cite{letaw-pfautsch,letaw}. 
This can be accomplished by a suitable addition and a subtraction in
the integrand, with the result 
\begin{equation}
{\dot F}_{\tau}(\omega)
=
-\frac{\omega}{2\pi}\Theta(-\omega)
+\frac{1}{2\pi^2}\int_0^{\infty}\textrm{d}s\cos (\omega s)
\left( 
\frac{1}{{(\Delta \mathsf{x})}^2} 
+\frac{1}{s^2}
\right) 
\ . 
\label{inermasacc}
\end{equation}
The first term in (\ref{inermasacc}) is the transition rate of a
detector in inertial motion, and the integral term is thus the
correction due to acceleration. As the correction is even in~$\omega$,
we see that 
the acceleration induces excitations and de-excitations with the same
probability. 

Note that the correction term in (\ref{inermasacc}) is nonvanishing
for
\emph{every\/} noninertial trajectory. Note also that inversion of the
cosine transform in (\ref{inermasacc}) shows that ${\dot
F}_{\tau}(\omega)$ fully determines ${(\Delta \mathsf{x})}^2$ as a
function of $s$ and~$\tau$.

From (\ref{inermasacc}) it follows that $\dot{F}_{\tau}(\omega)$ has a
large $|\omega|$ expansion that proceeds in inverse powers
of~$\omega^2$, with coefficients given by $\tau$-derivatives
of~$\mathsf{x}(\tau)$. In the leading order we obtain
\begin{equation}
\label{largeomega}
\dot{F}_{\tau}(\omega)
=
-\frac{\omega}{2\pi}\Theta(-\omega)
+\frac{\ddot{\mathsf{x}} \cdot {\mathsf{x}}^{(3)}}{24 \pi^2 \omega^2}
+O \! \left( \omega^{-4} \right) 
\quad\quad 
\mathrm{as}\,\,|\omega| \rightarrow\infty
\ , 
\end{equation}
which shows that for a generic trajectory the first correction to the
inertial response is of order~$\omega^{-2}$.

\section{Examples}
\label{sec:examples}

A case-by-case analysis of all stationary trajectories 
shows that 
the transition rate 
(\ref{resultado1-infty}) for them 
agrees with that obtained with the 
regularisation that relies on
stationarity~\cite{letaw-pfautsch,letaw}. 
In particular, in the special case of 
uniform acceleration of magnitude $a$ we have the
Planckian spectrum, 
\begin{equation}
\label{eq:Planck}
\dot{F}_{\tau}(\omega)
=
\frac{\omega}{2\pi}
\frac{1}{\mathrm{e}^{2\pi\omega/a}-1} 
\ . 
\end{equation}

As an example of nonstationary motion, consider a detector that moves
in a timelike plane with the proper acceleration 
$a/ (1 +
\mathrm{e}^{-a\tau} )$, where $a$ is a positive constant. 
In the distant past the trajectory is asymptotically 
inertial, and we obtain the transition rate 
\begin{equation}
{\dot F}_{\tau}(\omega)
=
-\frac{\omega}{2\pi}\Theta(-\omega)
+ 
O(\mathrm{e}^{2a\tau})
\ , 
\ \ \ 
\tau\to-\infty
\ , 
\end{equation}
where the $O$-term holds uniformly in~$\omega$. 
In the distant future the trajectory has 
asymptotically uniform acceleration of magnitude~$a$, 
and we obtain the transition rate
\begin{equation}
\label{eq:as-unruh-future}
\dot{F}_{\tau}(\omega)
=
\frac{\omega}{2\pi}
\frac{1}{\mathrm{e}^{2\pi\omega/a}-1}
+ o(1)
\ , 
\ \ \ 
\tau\to\infty
\ , 
\end{equation}
where $o(1)$ stands for a term that goes to zero as
$\tau\to\infty$. The first term in (\ref{eq:as-unruh-future}) is the
transition rate (\ref{eq:Planck}) in uniform acceleration. The
asymptotics thus agrees with what one would expect on physical
grounds, both in the future and in the past.

\section{Does shape matter?}
\label{sec:shape?}

The above results rely on the choice (\ref{lorentzian}) for the
profile function. While all sufficiently regular profile functions are
known to yield the same $\epsilon\to0_+$ limit for inertial
motion~\cite{Schlicht:thesis}, it is at present not known to what
extent the $\epsilon\to0_+$ limit might depend on the profile function
for more general motions. 

There is however a modified notion of spatial smearing in which we
have been able to establish a result on profile-independence. For
positive $\epsilon$, the transition rate with this modified smearing
reads
\begin{equation}
\label{eq:smearedFdot-def}
\dot{F}_{\tau}^{(\epsilon)} (\omega)
: =
\int_{\boldsymbol{\xi} \ne \boldsymbol{\xi}'}
\mathrm{d^3}\xi
\,
\mathrm{d^3}\xi' 
\; 
\epsilon^{-6} 
f(\boldsymbol{\xi}/\epsilon) 
\, 
f(\boldsymbol{\xi}'/\epsilon)
\, 
{G}_{\tau} 
(\boldsymbol{\xi} , \boldsymbol{\xi}'; \omega)
\ , 
\end{equation}
where 
\begin{equation}
\label{eq:G-def}
G_{\tau} 
(\boldsymbol{\xi} , \boldsymbol{\xi}' ; \omega)
:=  
2 \,
\mathrm{Re}\int_{0}^{\infty} \mathrm{d}s
\,\,
\mathrm{e}^{-i \omega s}
\, 
\langle 0\vert 
\phi \!\left( \mathsf{x}(\tau)  
+\xi^i\mathsf{e}_i(\tau) \right)
\phi \!\left( \mathsf{x}(\tau-s) 
+{{\xi'}{\vphantom{\xi}}^j} \mathsf{e}_j(\tau-s) \right)
\! 
\vert 0\rangle
\ . 
\end{equation}
Equations (\ref{eq:smearedFdot-def}) and (\ref{eq:G-def}) would follow
from (\ref{defexcitation-sharp-infty}) with the replacement
(\ref{eq:smeared-phi}) if it were known that 
the interchange of the $\mathrm{d}s$ 
and 
$\mathrm{d^3}\xi
\,
\mathrm{d^3}\xi'$ 
integrals is valid in a sense in which $G_{\tau} (\boldsymbol{\xi} ,
\boldsymbol{\xi}' ; \omega)$ (\ref{eq:G-def}) 
contains no distribution with support at $\boldsymbol{\xi} =
\boldsymbol{\xi}'$. While we do not know whether the interchange can
be justified in this sense, we shall take equations 
(\ref{eq:smearedFdot-def}) and (\ref{eq:G-def}) as a
\emph{definition\/} of a detector model in their own right, arguing
that this model captures at least some of the effects of the spatial
smearing of section~\ref{sec:profile}.

Now, if the trajectory is real analytic and satisfies suitable falloff
conditions in the distant past, and if the profile function $f$ is
smooth and has compact support, it can be shown \cite{louko-satz} that
$\dot{F}_{\tau}^{(\epsilon)} (\omega)$ is well defined by
(\ref{eq:smearedFdot-def}) and (\ref{eq:G-def}) for sufficiently
small~$\epsilon$, and the $\epsilon\to0_+$ limit exists and is given
by~(\ref{resultado1-infty}). As this limit agrees with that obtained
with the Lorentzian profile function (\ref{lorentzian}) (which is not
of compact support), we suspect that the equivalence of the
two models of spatial smearing could be established for at least some
classes of profile functions.

\section{Discussion}
\label{sec:discussion}

We have shown that regularising the transition rate of an accelerated
Unruh-DeWitt detector on Minkowski space by a spatial profile is a
mathematically well-defined procedure and yields physically viable
predictions in a number of situations.  For the Lorentz-function
spatial profile (\ref{lorentzian}) the zero size limit could be
computed explicitly, leading to the transition
rate~(\ref{resultado1-infty}). For other spatial profiles the results
remain to some extent inconclusive but they suggest that the zero-size
limit may not be sensitive to the details of the profile.

We re-emphasise that the need for a spatial smearing arose because we
chose to address the instantaneous transition rate 
\emph{while the interaction continues to be switched on\/}, 
rather than the total excitation probability after the interaction has
been smoothly switched on and off by an external agent. It would be of
interest to examine in comparison a pointlike detector whose smooth
switching function is allowed to approach the step-function: Might there
exist limiting prescriptions that reproduce the effects of spatial
smearing?

If the detector is turned on sharply at the finite proper time~$\tau_0$, 
the transition rate formula (\ref{resultado1-infty}) is replaced by 
\cite{louko-satz}
\begin{equation}
\label{resultado1-conc}
\dot{F}_{\tau}(\omega)
=
-\frac{\omega}{4\pi}+\frac{1}{2\pi^2}
\int_0^{\tau - \tau_0}
\textrm{d}s
\left( 
\frac{\cos (\omega s)}{{(\Delta \mathsf{x})}^2} 
+ 
\frac{1}{s^2} 
\right) 
\ \ +\frac{1}{2\pi^2 (\tau - \tau_0)}
\ , 
\ \ \tau> \tau_0
\ , 
\end{equation}
which is asymptotically proportional to ${(\tau-\tau_0)}^{-1}$ as
$\tau\to\tau_0$. The total transition probability, obtained by
integrating the transition rate~(\ref{resultado1-conc}), is therefore
infinite, owing to the violent switch-on event, regardless how small
the coupling constant in the interaction Hamiltonian is. For the
stationary trajectories the transition rate (\ref{resultado1-infty})
of a detector switched on in the asymptotic past is constant in time,
and the total transition probability is again infinite, now owing to
the infinite amount of time elapsed in the past. In these situations
one may therefore have reason to view our results, all of which were
obtained within first-order perturbation theory, as suspect. However,
in situations where the detector is switched on in the asymptotic past
of infinite proper time and the total probability of excitation
($\omega>0$) is finite, the first-order perturbation theory result
should be reliable at least for the excitation rate, although the
total probability of de-excitation ($\omega<0$) then still
diverges. This situation occurs for the asymptotically inertial
trajectory discussed in section~\ref{sec:examples}, and we expect it
to occur whenever the proper acceleration vanishes sufficiently fast
in the distant past.

It would be interesting to investigate to what extent our results can
be generalised to the variety of situations to which Schlicht's
Lorentzian profile detector was generalised in
\cite{Langlois-thesis,Langlois}. For example, do the formulas
(\ref{resultado1-infty}) and 
(\ref{resultado1-conc})
generalise to
spacetime dimensions other than four, and if yes, what is the form of
the subtraction term? Does the clean separation of the spectrum into its
even and odd parts continue? 
Further, to what extent can the notion of spatial profile be employed 
to regularise the transition rate in a curved spacetime, 
presumably reproducing known results for stationary trajectories 
\cite{sonego-westman} but also allowing nonstationary motion? 
In particular, might there be a connection with the
regularisation prescriptions of the classical self-force
problem~\cite{DetWhit,poisson-livrev,poisson-gr17,anderson-wiseman}? 
Finally, would a nonperturbative treatment be
feasible?

\ack
JL thanks the organisers of the NEB XII meeting for the invitation to
present this work and their kind hospitality, Pierre Martinetti for
discussions during the meeting and The British Council for travel
support. JL acknowledges the hospitality and financial support of the
Isaac Newton Institute programme ``Global Problems in Mathematical
Relativity" and the Perimeter Institute for Theoretical Physics and
thanks Howard J. Magnuson for hospitality during the preparation of
the manuscript. AS was supported by an EPSRC Dorothy Hodgkin Research
Award to the University of Nottingham.

\end{document}